\documentclass[10pt]{article}
\usepackage[OE]{express}
\usepackage{fixltx2e}
\usepackage{mathrsfs}

\newcommand{\Fi}{\mathscr{F}}

\begin{document}

\title{Dual-wavelength fiber Fabry-Perot cavities with engineered birefringence}

\author{S\'{e}bastien Garcia,\authormark{1} Francesco Ferri,\authormark{1} Konstantin Ott,\authormark{1,2} Jakob Reichel,\authormark{1} and Romain Long \authormark{1,*}}

\address{\authormark{1}Laboratoire Kastler Brossel, ENS--Universit\'e
  PSL, CNRS, Sorbonne Universit\'e, Coll\`ege de France,
  24 rue Lhomond, 75005 Paris, France\\
%\authormark{2}Current address: Department of Physics, ETH Z\"urich, CH-8093 Z\"urich, Switzerland\\
\authormark{2}LNE-SYRTE, Observatoire de Paris--Universit\'e PSL, CNRS, 
   Sorbonne Universit\'e, 61 Avenue de l'Observatoire, 75014 Paris, France}

\email{\authormark{*}long@lkb.ens.fr} %% email address is required


\begin{abstract}
  We present a method to engineer the frequency splitting of polarization eigenmodes in fiber Fabry-Perot (FFP) cavities.  Using specific pattern of multiple CO$_2$ laser pulses, we machine paraboloidal micromirrors with controlled elliptical shape in a large range of radii of curvature. This method is versatile and can be used to produce cavities with maximized or near-zero polarization mode splitting.  In addition, we realize dual-wavelength FFP cavities with finesse exceeding $40\,000$ at 780$\,$nm and at 1559$\,$nm in the telecom range. We provide direct evidence that the birefringent frequency splitting in FFP cavities is governed only by the geometrical shape of the mirrors, and that the astigmatism of the cavity modes needs to be taken into account for specific cavities.
\end{abstract}


%\bibliographystyle{osajnl}
%\bibliography{tailored_birefringence_OE}

%%%%%%%%%%%%%%%%%%%%%%%%%%  body  %%%%%%%%%%%%%%%%%%%%%%%%%%
\section{Introduction}

Since their first realization \cite{Colombe2007,Hunger2010},
high-finesse Fiber Fabry-Perot (FFP) cavities with laser-machined
mirrors have rapidly spread and are now routinely used in many
experiments, enabling miniature setups with stable, enhanced
coupling between light and different material systems. They have been
combined with atomic systems (cold atoms \cite{Colombe2007,
  Barontini2015, Uphoff2015, Gallego2016}, ions
\cite{Brandstatter2013, Steiner2013}, or molecules
\cite{Toninelli2010}), quantum dots and quantum wells
\cite{Muller2010, Miguel-Sanchez2013}, nitrogen vacancy centers
\cite{Albrecht2013, Kaupp2016}, carbon nanotubes \cite{Jeantet2016,
  Hummer2016}, as well as opto-mechanical devices
\cite{Flowers-Jacobs2012, Kashkanova2016, Zhong2017}. Similar
laser-machined mirrors on free-space substrates \cite{Barbour2011} are
also being used successfully with solid-state emitters.  To progress
on the development of these miniature light-matter interfaces, efforts
have been devoted to improving the production of the laser-machined
fiber micromirrors
\cite{Petrak2011,Hunger2012,Greuther2014,Ott2016}. The
shape carved into the fiber end-facets and its surface quality are
crucial parameters for the cavity mode volume and for the
finesse of the resonator, determining the strength of the light-matter
coupling. Another property that needs to be fully controlled
%to get an optimized and well-defined coupling: 
is the polarization of the cavity modes. 
%The reflection phase shift of optical mirrors generally depends on the
%polarization of the incident light field, resulting in a birefringent
%splitting of the resonance frequencies in Fabry-Perot cavities. 
Cavity birefringence is an important factor not only for cavity
quantum electrodynamics (CQED), but indeed for most applications of
high-finesse cavities, ranging from precision spectroscopy to
gravitational wave detection and measurement of vacuum magnetic
birefringence \cite{Bielsa2009,Fleisher2016}. 
It has been shown in
\cite{Uphoff2015} that the birefringence of FFP cavities is mainly
due to the geometrical asymmetry of the shape of the mirrors. This
effect is particularly important for FFP cavities because the radii of
curvature of the fiber mirrors are much smaller than of typical
macroscopic ones, and because the control of the production of the
mirrors is challenging due to the non-linearity of the CO$_2$ laser
ablation process.

Depending on the application, this birefringence can be a problem or
an advantage. 
%In CQED experiments, a large
%birefringent splitting can be used to guarantee a well-defined
%polarization of the intracavity field. On the other hand, it precludes
%the use of circularly polarized light.
%  For some CQED experiments, a
% birefringence mode-splitting represents a drawback. For example, 
Polarization-degenerate cavities can be used to address closed atomic 
transitions and to implement photonic polarization qubits
\cite{Stute2012, Reiserer2014}.
%  Uphoff2016}. Besides, such cavities can be probed by circularly
%polarized light allowing the 
%, providing a much better approximation of an ideal two-level system.  
To realize FFP cavities with very low
birefringence, one method is to use imperfectly shaped fiber mirrors
and compensate the birefringence of each mirror by rotating one fiber
with respect to the other \cite{Uphoff2015}. Another
%, more deterministic technique to
%obtain FFP cavities with degenerate polarization modes, 
technique is to produce
micromirrors with a high-degree of rotational symmetry by rotating the
fiber during the ablation process \cite{Takahashi2014}.

However, the frequency splitting of eigenmodes can also be used as a
resource. For example, if one polarization mode of the cavity is
closely resonant to the transition of an emitter, a large splitting
detunes the second, orthogonally polarizated mode far away
from this transition, minimizing its detrimental role \cite{Gehr2010,
  Volz2011} as an extra decay channel. It also enables the use of the
cavity as a polarization filter, and makes it possible to prepare an
intracavity field with well-defined polarization by appropriately
tuning the cavity.
% The ability to prepare a single
% mode with a well-defined polarization can also be an advantage as the
% cavity acts as a polarization filter and can avoid a tedious
% polarization control of the incoupling light.
Such a controlled
splitting could even be used to address two different transitions of
an optical emitter with the two splitted polarization eigenmodes.

FFP cavities with large birefringent frequency splitting can only be obtained by producing fiber mirrors with strongly asymmetrical
profiles. Recently, we introduced a new method of machining fibers
based on a laser ablation approach using spatial pattern of multiple
pulses \cite{Ott2016}. It has allowed us to realize FFP cavities with
millimeter length scale. In this article, we leverage this method to
produce strongly elliptic paraboloidal fiber mirrors, and measure a
large birefringent splitting. To compare different production methods, we introduce the "geometrical birefringence" of a cavity, which is a parameter independent of the wavelength of the light and of the finesse of the resonator.  We determine theoretically the
orientation of the polarization eigenaxes of the resonator for a
non-degenerate cavity. We also produce fiber mirrors with a very low
asymmetry, ideally suited for cavity with low birefringence, by
compensating the imperfection of the standard single pulse laser
ablation. This method, which does not require rotating the fiber, is
therefore applicable to any substrate. In addition, we report in
detail the operation of a FFP cavity working at the two wavelengths of
1559$\,$nm and 780$\,$nm with finesse exceeding $40\,000$. We measure
the variation of the finesse with the cavity length for both
wavelengths and show the effect of the asymmetry of the cavity mode.
We also measure the birefringent frequency splitting of the cavity for the two wavelengths and provide direct evidence that this splitting is governed by its geometrical birefringence for a given finesse and wavelength.

\section{Birefringence of FFP cavities}
	
\subsection{Geometrical birefringence of a cavity} 

In high-finesse cavities, and especially in FFP cavities, a frequency
splitting is usually observed between the two polarization eigenmodes
of a given transverse and longitudinal cavity mode.  In a cavity with
isotropic mirror coatings, the birefringent frequency splitting stems
from the rotational asymmetry of the mirror. Indeed, for light of
wavelength $\lambda$, the birefringent dephasing $\phi_{\mathrm{m}}$
between the two polarization modes due to a single reflection on an
elliptic paraboloidal mirror is given by \cite{Uphoff2015}:
\begin{equation}
\phi_{\mathrm{m}} = \frac{\lambda}{2 \pi} \delta_{\mathrm{m}} \ \ \ \ \mathrm{with} \ \ \ \ \delta_{\mathrm{m}}  = \frac{1}{R_{a}} - \frac{1}{R_{b}}=-\frac{\gamma}{R_b}\ \ \ \ \mathrm{where} \ \ \ \
\gamma=\frac{R_a-R_b}{R_a}
\label{eq:phasebiref}
\end{equation}
In this expression, we define the parameter $\delta_{\mathrm{m}}$ as the geometrical birefringence of the mirror. It only depends on the long radius of curvature (ROC) $R_{a}$ and on the short one $R_{b}$ along the eigenaxes of the mirror. It represents the intrinsic effect of the mirror shape and can be controlled by the relative difference $\gamma$ of the ROCs, which is equal to the excentricity squared of the elliptical cross-section of the mirror parallel to the fiber endfacet. As $\phi_{\mathrm{m}}$ depends on the ratio $\lambda / R_b$, controlling the birefringent dephasing becomes harder for small ROCs or large wavelengths. These equations are not limited to optical mirrors but apply to any mirrors in the full electromagnetic spectrum.   

In the following, we consider two birefringent mirrors $M_1$ and $M_2$ with isotropic coatings and geometrical birefringences given respectively by  $\delta_1$ and $\delta_2$. By convention, the fast axis of $M_1$ is taken along the vertical direction and the fast axis of $M_2$ is rotated by an angle $\theta$ with respect to the one of $M_1$ (see Fig.~\ref{fig:IntraPolRot}(a)). When the mirrors are birefringent, the polarization eigenmodes inside the cavity can be calculated via the Jones matrix formalism \cite{jones1941new}. As the ROCs in usual FFP cavities are much larger than the wavelengths, birefringent dephasing is small: $\phi_{\mathrm{m}} \ll 1$. In the small phase approximation, the polarization eigenmodes in the cavity are linear and the ratio $\epsilon$ of the birefringent frequency splitting $\Delta \nu$ to the resonator linewidth $\kappa$ (defined as Half Width Half Maximum in frequency) can be expressed as : 

\begin{equation}
\epsilon\left( \theta \right) = \frac{\Delta \nu \left( \theta \right)}{\kappa} =  \frac{\lambda \Fi }{2 \pi^2} \Delta_c \left( \theta \right) \ \ \ \ \mathrm{with} \ \ \ \ \Delta_\mathrm{c} \left( \theta \right)= \sqrt{\delta_1^2 + \delta_2^2 + 2 \delta_1 \delta_2 \cos\left( 2 \theta \right) }  
\label{eq:splitting}
\end{equation}
where $\Fi$ is the finesse of the cavity and where we define
the geometrical birefringence of the cavity, $\Delta_\mathrm{c}$. The latter
depends only on the shape of the two mirrors via $\delta_{\mathrm{1}}$
and $\delta_{\mathrm{2}}$, and on their relative orientation
$\theta$. $\Delta_\mathrm{c}$ results only from the geometrical
properties of the cavity while the resonant wavelength $\lambda$ and
the finesse $\Fi$ featuring in $\epsilon$ depend only on the
properties of the dielectric coating.

For applications where a given birefringence frequency splitting is needed, it can be adjusted by the angle $\theta$ between the two fast axes of the birefringent mirrors in the range $|\delta_1 - \delta_2| < \Delta_\mathrm{c} \left( \theta \right) <  \delta_1 + \delta_2$. A perfect degenerate-mode cavity can only be obtained with mirrors that have exactly the same geometrical birefringence. However, variations in the shape productions lead to non-perfect cancellations of the cavity geometrical birefringence. A reliable solution to approach degeneracy is to use mirrors that have an excellent rotational symmetry and inherently a very low geometric birefringence. On the contrary, if a specific non-zero splitting is targeted, the best solution is to produce mirrors with approximately the same geometrical birefringence, slightly above one half of the targeted cavity one. The cavity birefringence will then be easily adjusted by a small rotation of one mirror remaining close to $\theta=0$ and thus being less sensitive to angle adjustment errors. If a precise absolute frequency splitting $\Delta \nu$ is needed, the wavelength and the length $L$ of the  cavity must be taken into account to design the geometrical birefringence of the cavity according to the relation 
\begin{equation}
\Delta \nu \left( \theta \right) = \frac{c \lambda}{8 \pi^2 L} \Delta_\mathrm{c} \left( \theta \right),  
\label{eq:splitting2}
\end{equation}
where $c$ is the speed of light.

\subsection{Polarization of the cavity mode for birefringent cavities}

\begin{figure}[htb]
	\centering\includegraphics[width=0.6\columnwidth]{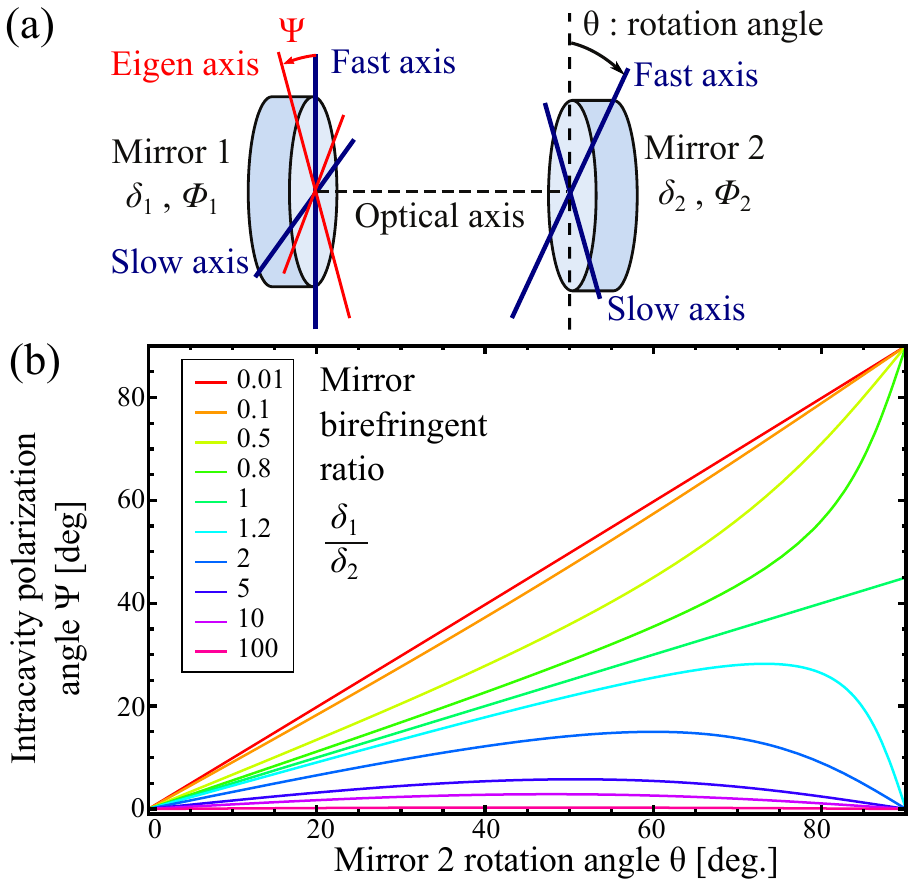}
	\caption{(a): Scheme of a cavity with birefringent mirrors
          showing the mirrors and the intracavity eigenbases. $\theta$
          is the angle between the fast axes of the two
          mirrors. $\Psi$ is the angle between the fast eigenmode of
          the cavity and the fast axis of mirror 1. (b): The
          intracavity polarization eigenbasis is rotated by an angle
          $\Psi$ relative to the first mirror's birefringent basis and
          depends only on the relative rotation angle $\theta$ of the
          birefringent basis of the second mirror and on the ratio of
          the geometrical birefringences
          $\frac{\delta_1}{\delta_2}$. }
	\label{fig:IntraPolRot}
\end{figure}

In an experiment with a non-degenerate cavity, the orientation of the polarization eigenmodes with respect to the complete apparatus can be of crucial importance. For example in atomic systems, it is required to achieve precise polarization dependent transitions for optical pumping. 

If we consider the same setting as previously with two birefringent mirrors $M_1$ and $M_2$ with isotropic coatings and geometrical birefringence  $\delta_1$ and $\delta_2$, then the fast eigenmode of the cavity (i.e. the highest frequency mode) will have an angle $\Psi\left(\theta \right) $  relative to the fast axis of $M_1$ given by:
\begin{equation}
\Psi\left(\theta \right) = \arctan \left[\frac{1}{\sin \left(2 \theta \right)} \left(\sqrt{1 + \left(\frac{\delta_1}{\delta_2}\right)^2 + 2\frac{\delta_1}{\delta_2} \cos \left(2 \theta \right)}-\frac{\delta_1}{\delta_2} - \cos \left(2 \theta \right)  \right) \right]
\label{eq:intraPol}
\end{equation}

The value of the angle $\Psi$ depends only on the ratio of the geometrical birefringence of the mirrors $\frac{\delta_1}{\delta_2}$ and on the angle $\theta$ between the fast axes of the two mirrors. The evolution of the angle of the polarization eigenmode basis is shown on Fig.~\ref{fig:IntraPolRot}(b). If the geometrical birefringence of one mirror is much larger than of the other one $\left(\frac{\delta_2}{\delta_1} \ll 1 ~~\mathrm{or}~~ \frac{\delta_2}{\delta_1} \gg 1 \right)$, then the eigenmode basis is aligned with the fast axis of the strongest birefringent mirror. When both are equal $\delta_2 = \delta_1$, the fast eigenmode is on the bisector of the fast axes of the two mirrors.

\subsection{Spatial distribution of astigmatic cavity modes}

\begin{figure}[htb]
	\centering\includegraphics[width=0.8\columnwidth]{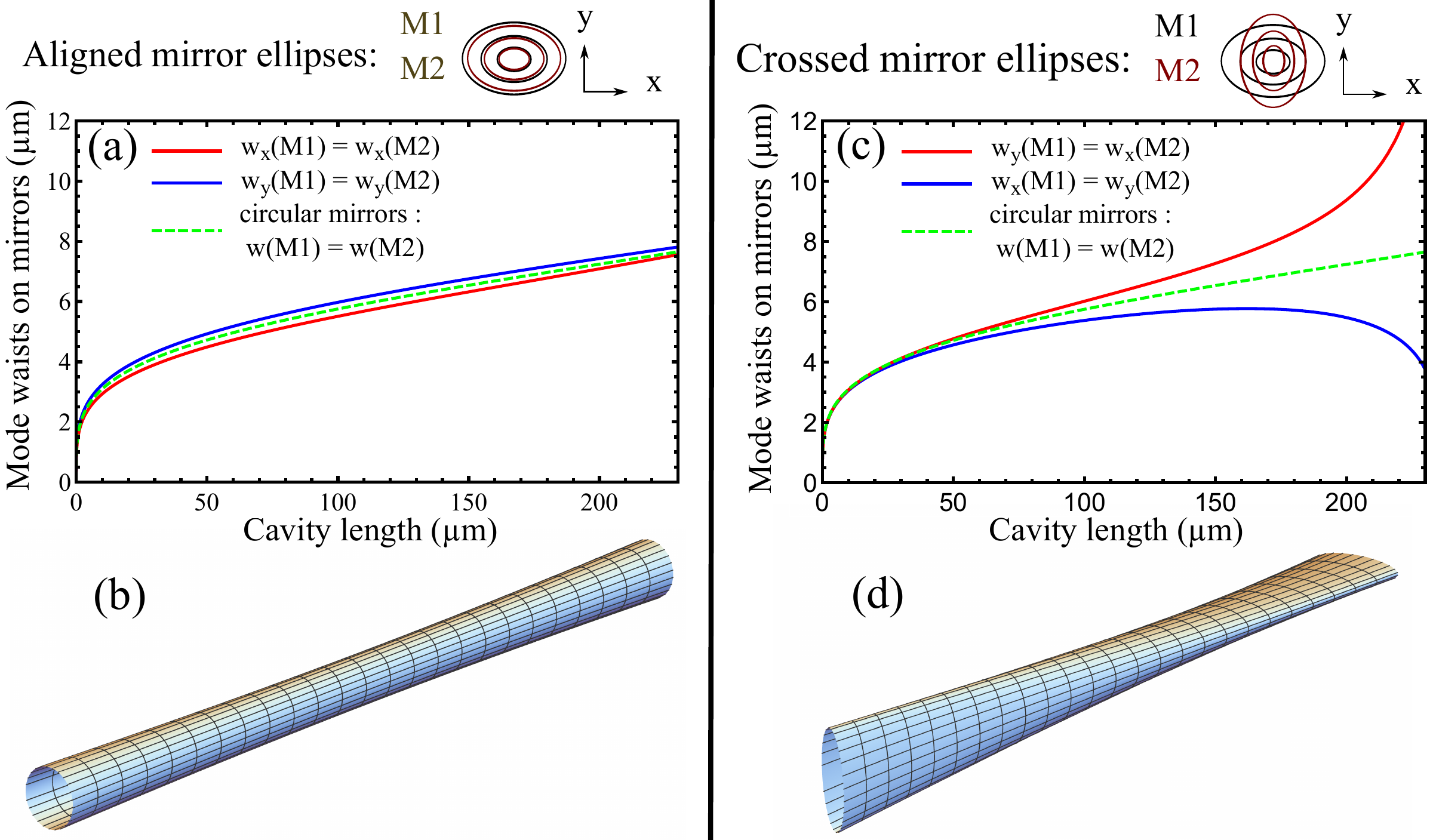}
	\caption{Comparison of the spatial distributions of
          the intracavity mode when the two fast axes are
          parallel (left) or orthogonal (right). The birefringent
          cavity is formed by two identical elliptic paraboloidal mirrors with
          ROCs of 235 $\,\mu$m and 355 $\,\mu$m. We plot the mode size on the two
          mirrors along the two principal axes (a) and (c). We also
          plot the mode size on the two mirrors for a non-astigmatic
          290$\,\mu$m ROC circular cavity. (b) and (d): contour of
          equal intensity for a cavity length
          of 230 $\,\mu$m.}
	\label{fig:spatial_distribution}
\end{figure}

For applications where a maximal frequency splitting between the polarization eigenmodes is required,
mirrors with strong geometrical birefringence are needed and the two
fast axes have to be close to parallel. This implies to produce
mirrors with very different radii along the two eigenaxes, noted $x$
and $y$, which are then highly astigmatic. This affects the spatial
distribution of the cavity mode. For the simple case of two identical
mirrors with ROCs $R_x$ and $R_y$ and with parallel fast axes, the
waists are located in the center of the cavity for both axes, but have
a different value $w_{0,x}$ and $w_{0,y}$ for each one. The mode is
then elliptical and the ratio of the waists for a cavity length $L$ is
given by
$\displaystyle \frac{w_{0,x}}{w_{0,y}}=\left(\frac{2 R_x-L}{2
    R_y-L}\right)^{1/4}$. If the length of the cavity $L$ is not too
close to twice the smallest ROC, this ratio scales slowly with $L$ and
stays roughly close to one. We are then close to the case of a
circular cavity formed by two identical mirrors with a ROC
corresponding to the average of $R_x$ and $R_y$ (see Fig.~\ref
{fig:spatial_distribution}).

If instead a degenerate-mode cavity is targeted, a crossed
configuration where the two fast axes of each mirror are orthogonal
will minimize the cavity birefringence by compensating the geometrical
one of each mirror. For two identical mirrors, the waists have the
same value for both directions but their positions for each direction
lie on opposite sides of the cavity center at a distance given by
$\displaystyle \frac{L}{2} \frac{\vert R_x-R_y \vert}{2 L-
  R_x-R_y}$. When the resonator length tends toward the smallest ROC,
the cavity gets closer to the limit of the stability region. The
waists for each direction lie on the two opposite mirrors (on the
mirror having the largest ROC along this direction) and the mode size
on the facing mirror starts to diverge (see Fig.~\ref
{fig:spatial_distribution}). This induces clipping losses that limit
the finesse.

In the configuration where the fast axes of the two mirrors are neither parallel nor orthogonal, the cavity is twisted, featuring general astigmatism with a rotating elliptic intensity distribution along the resonator axis \cite{Arnaud1969,Habraken2007, Weber2012}.

Beyond polarization control, our ability to produce mirrors with no rotational symmetry can be leveraged to tailor the spatial shape of cavity modes, for example to compensate the astigmatism of some ring cavity types.

\section{Versatile CO$_2$ laser dot machining of FFP cavity micromirrors}

	\subsection{CO$_2$ laser dot machining} 
	
	\begin{figure}[p]
	\centering\includegraphics[width=0.95\columnwidth]{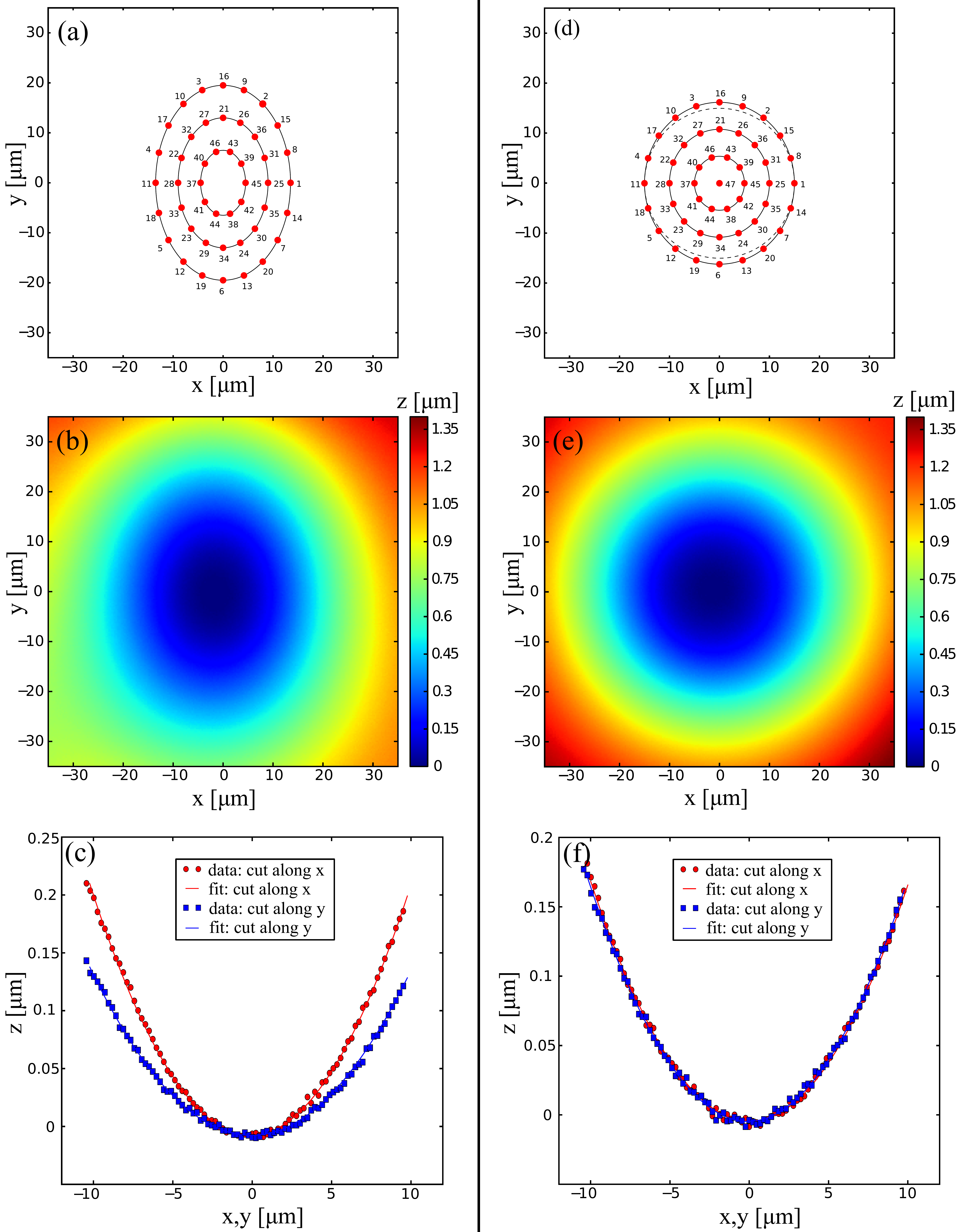}
	\caption{Realization of elliptic (left) and circular (right) fiber mirror shapes. (a,d): Shot patterns used to mill the fiber endfacet with CO$_2$ laser pulses. The red dots indicate the relative position of the shots centered on the fiber core, and the numbers the corresponding number in the pulse series. The black curves mark the alignment ellipses. The dashed black circle on (d) is represented to emphasize the slight ellipticity of the shot pattern that optimizes the symmetry of the carved shape (cf. text). (b,e): Resulting profiles of the fiber shapes after laser ablation, showing large (b) and small (e) geometrical birefringences. (c,f): Cuts of the fiber profile data and corresponding paraboloidal fits along the eigenaxes, $x$ and $y$, of the shot patterns, centered on the minimum value to emphasize (a)symmetry. The fitted radii of curvature of these structures along their principal axes are $355\,\mu$m and $236\,\mu$m for the elliptic paraboloid mirror and $294\,\mu$m and $289\,\mu$m for the almost circular mirror. }
	\label{fig:ellip_circ_combined}
\end{figure}

We have recently developed a CO$_2$ laser ablation setup, described in
detail in~\cite{Ott2016}, where the position of the fiber (or
substrate) is controlled in all directions by a combination of
state-of-the-art translation stages. This setup features several
substantial improvements such as the automation of fiber alignment,
in-situ phase-shifting profilometry and fast modification of the laser
ablation beam diameter. Most importantly, it gives us the ability to
carve structures into the fiber endfacets by using multiple laser
pulses ("shots") precisely positioned relative to the fiber core
center and with adjustable pulse lengths. This has allowed us to
produce large circular fiber mirrors with large ROC, mandatory to
realize FFP cavities longer than a millimeter~\cite{Ott2016}.
	
Moreover, this technique extends the range of realizable surface
structures well beyond the circular symmetry while preserving the very
low surface roughness required to sustain high finesse cavity modes
and obtained thanks to surface tension smoothing
\cite{Hunger2012}. The relative precision of the transverse motion of
the substrate by the translation stages is below $100\,$nm, thus
offering unprecedented control over the shape of the carved structure
via the choice of an adequate geometry of the shot pattern. In this
article, we focus on the control of the elliptic shape of mirrors (and
thus on the control of their birefringence) by using concentric
elliptical shot patterns as shown on
Fig.~\ref{fig:ellip_circ_combined}(a).
	
	As the laser ablation is a highly non-linear process, finding a suitable shot pattern to produce elliptic paraboloidal structures needs at first some empirical investigations before a systematic optimization can be done. We mention here some general rules to converge towards such a pattern. First, neighboring shots should not be done consecutively to maintain a homogeneous heating of the substrate during the laser ablation process. As can be seen on Fig.~\ref{fig:ellip_circ_combined}(a), consecutive shots are separated by 3 positions in our optimized elliptical pattern. Second, for the same reason, we set a waiting time of $1$ second between two pulses. Third, the most centered shot should be done last, specially because the smoothing effect by surface tension occurs over a melted region of the mirror, which is much larger than the zone where single shot ablation occurs.

	\subsection{Elliptic paraboloidal micromirrors} 
	
	Strongly elliptic paraboloidal structures were obtained by
        using shot patterns with a large ellipse radii ratio $s_e$
        (typically close to 2), as the one presented on
        Fig.~\ref{fig:ellip_circ_combined}(a) that contains 46
        shots. The shape shown on
        Fig.~\ref{fig:ellip_circ_combined}(b) is obtained by CO$_2$
        laser ablation of a multimode fiber (Cu50, IVG Fiber Ltd). We
        use a CO$_2$ laser beam with a waist of 78$\,\mu$m, a power of
        750$\,$ mW, and single shot pulse length of 14.7$\,$ms. We
        extract the two ROCs ($R_a=355 \,\mu$m and $R_b=236 \,\mu$m )
        along the eigenaxes of the mirrors by using a bi-dimensional
        fit of the structure within a 10$\,\mu$m radius around the
        center. By taking into account the measured $1.4\,$nm standard
        deviation of the profilometry noise, the standard deviations
        of the fitted ROCs are much lower than $1\,\mu$m. The
        eigenaxes of the mirror are aligned with the eigenaxes,
        noted $x$ and $y$, of the elliptical shot patterns. The cuts
        of the profile and of the bi-dimensional fit along these axes
        are presented on Fig.~\ref{fig:ellip_circ_combined}(c). The
        produced shape has mean absolute deviations from an elliptic
        paraboloid as small as $5\,$nm over a circular region of
        interest which has a diameter of 30 $\mu$m.
	
In order to tune the value of the mirrors geometric birefringences of different fiber types, we vary the single shot pulse length (between $14-21.5\,$ms), the ratio of the shooting ellipse radii (over the range $1.44-2.15$) and the homothetic scaling of the shot positions. The mirror ROCs along the two eigenaxes were mostly designed to be in the range of respectively   $R_a \sim 350 \,\mu$m  and  $R_b \sim 250 \,\mu$m   and show correlated fluctuations from one fiber to another of about $\pm 5 \%$. By changing the pulse length and the scaling, we have tuned them from $R_a \simeq 250 \,\mu$m  and  $R_b \simeq 170 \,\mu$m  up to $R_a \simeq 780 \,\mu$m  and  $R_b \simeq 350 \,\mu$m. 
This range was chosen according to the targeted characteristics of the cavity for our experiment which take into account the  birefringent frequency splitting but also the mode waist and mode volume, the cavity stability and the mode coupling to the fibers modes. In particular, we aimed at stable cavities with length of about 100$\, \mu$m and with relatively small mode volumes as required for CQED experiments. However, for other cavity geometries, the rotational asymmetry of the shape, and thus the geometric birefringence can be increased, if a large frequency splitting is needed.

	\subsection{Circular paraboloidal micromirrors}

\begin{figure}[htb]
	\centering\includegraphics[width=0.55\columnwidth]{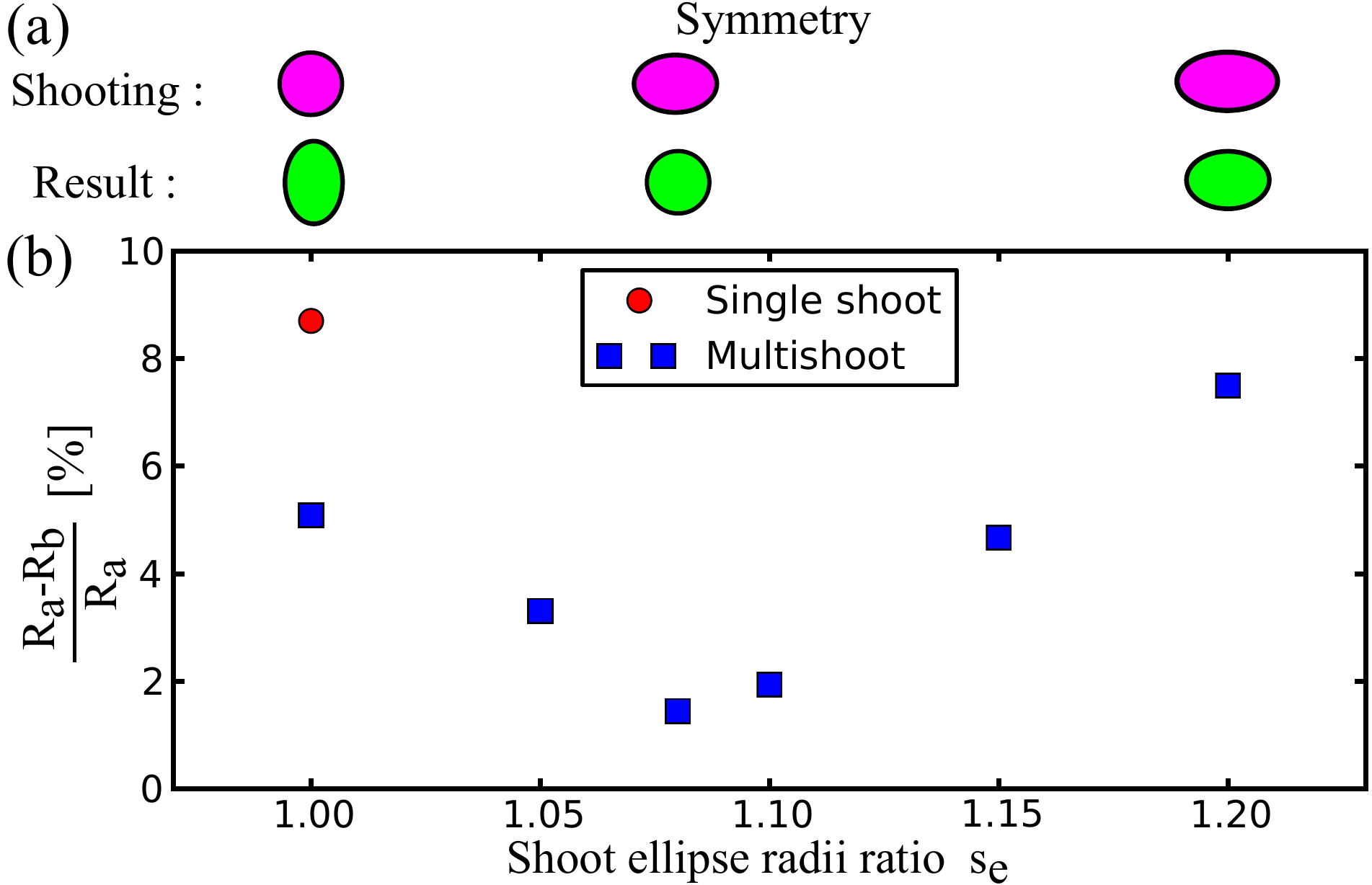}
	\caption{Compensation of CO$_2$ beam asymmetry with multi-shot technique. Each point is averaged over 3 shot fibers.  (a) Principle of the compensation : an elliptical shooting pattern in the longitudinal direction compensates the vertically elongated single shot result.  (b) Relative difference  $\gamma=\frac{R_a-R_b}{R_a}$ of the produced mirrors (Eigenaxes ROCs $R_a$, $R_b$)  as function of the shot pattern ellipticity $s_e$: ratio of major (Y) radius by minor (X) radius. Asymmetry is compensated down to $1.5$\% .}
	\label{fig:compensatedEllip}
\end{figure}

If nearly degenerate-mode cavities are targeted, mirrors with identical geometrical birefringence are required. To avoid extra alignment steps of mirror rotation and strongly asymmetric spatial distribution of the cavity mode, the best solution is to produce mirrors with very low geometric birefringence.
However, the mirror shapes produced with a single shot technique usually show some residual asymmetry that stems from imperfections of the CO$_2$ laser beam (non ideal circular polarized light, astigmatism). To eliminate this effect due to the lack of symmetry of the laser beam, one method is to rotate the fiber during the ablation process \cite{Takahashi2014}. Here, we compensate for it with the CO$_2$ dot machining method. We use a shot pattern with an ellipticity which is opposite to the ellipticity of the mirror's cross-section obtained by a single shot as shown on Fig.~\ref{fig:compensatedEllip}(a). The processed fiber is Thorlabs 980HP,  the CO$_2$ beam waist is 78$\,\mu$m, the produced ROCs are around 290$\, \mu$m, and the shot pattern is the same as the one of Fig.~\ref{fig:ellip_circ_combined}(d)). 

The relative difference  $\gamma=\frac{R_a-R_b}{R_a}$ of the ROCs of the two eigenaxes ROC when varying the pattern ellipticity are presented on Fig.~\ref{fig:compensatedEllip}(b). It is directly related to the geometrical birefringence $\delta_{\mathrm{i}}$  by a factor $\frac{1}{R_b}$ which remains approximately constant for a given cavity design.

	When we use the single shot technique, we obtain  mirrors with  $\gamma \simeq 9$\%, where we  have chosen the major axis to be vertical. The dot machining method with a circular pattern ($s_e = 1$) improves $\gamma$ due to the additional symmetry of the shooting pattern, but still presents a difference of about $5$\%. By using an elliptical shooting configuration whose major axis is orthogonal to the previous one (\textit{i.e.} horizontal), we can reduce the asymmetry down to about $1.5$\% between the eigenaxes ROCs on average over several fibers. Some fibers in the set show a relative difference below $1$\%, as the one presented in Fig.~\ref{fig:ellip_circ_combined}(e,f), paving the way for a reliable realization of degenerated polarization modes in FFP cavities. Moreover, contrary to the technique that relies on fiber rotation, this method is easily applicable to any substrate.

\section{Dual-wavelength operation of high-finesse FFP cavities}	\label{finesse}

After producing the mirror structures, the fibers have been coated with a dual-wavelength high-reflective coating for 780$\,$nm and 1559$\,$nm (ion beam sputtering technique performed by Laseroptik GmbH). The choice of this coating is motivated by the specific application of our fiber cavities in the domain of CQED with neutral \textsuperscript{87}Rb atoms. In our experiment, the TEM\textsubscript{00} cavity mode at 780$\,$nm  probes the \textsuperscript{87}Rb atoms on the D$2$ line, while the fundamental mode at 1559 $\,$nm acts as far-off-resonance optical lattice to trap the atoms along the cavity axis, achieving identical and maximal coupling to the probe. 
%We have built and tested two types of fiber cavities and measured the cavity finesse at both wavelengths for different cavity length.  

\subsection{Two types of FFP cavities}

To inject the laser light in the fiber cavities, we use single-mode (SM) fibers in order to get a stable coupling, which is one of the key advantages of FFP cavities. The coupling efficiency results then from the overlap between the out-coming mode of the fiber (approximately Gaussian) and the cavity mode. To optimize the coupling and finesse, one fiber is mounted on a 3-axis translation stage with piezo actuators, while the second fiber is mounted on a 3-axis rotation stage. We have tested two different types of FFP cavities:
\begin{enumerate}
\item SM-SM cavities: the 780$\,$nm laser light is coupled into one
  fiber (IVG Cu800, SM from 770$\,$nm to 1000$\,$nm), while the
  1559$\,$nm light is coupled into the other one (IVG Cu1300, SM in the range 1250-1600$\,$nm). These are commercial copper-coated fibers compatible with  ultra-high vacuum applications. The light transmitted through the cavity at 780$\,$nm can propagate efficiently in the SM fiber for 1559$\,$nm, while the transmission at 1559$\,$nm cannot be measured due to strong losses in the fiber for 780$\,$nm.

\item Photonic crystal (PC)-Multimode (MM) cavities:  both 780$\,$nm
  and 1559$\,$nm lights are injected through an "endlessly
  single-mode" photonic crystal fiber with large mode area (LMA-10
  from NKT Photonics). This fiber is specified for single-mode
  operation over a large wavelength interval. To be able to use this
  fiber in ultra-high vacuum, we replace the original acrylate coating
  with a home-made polyimide coating. The transmitted light is
  collected through the graded-index MM fiber (IVG Cu50) with essentially 100\% efficiency for both wavelengths. The large mode diameter of the PC fibers that we choose allows us to achieve a better fiber-to-cavity coupling compared to the SM-SM configuration. PCF-MM cavities are also less sensitive to misalignment and centering imperfections than the SM-SM ones, due to the large acceptance of the MM fiber. 
\end{enumerate}

\subsection{Finesse at 780$\,$nm and 1559$\,$nm}

\begin{figure}[htb]
	\centering\includegraphics[width=0.9\columnwidth]{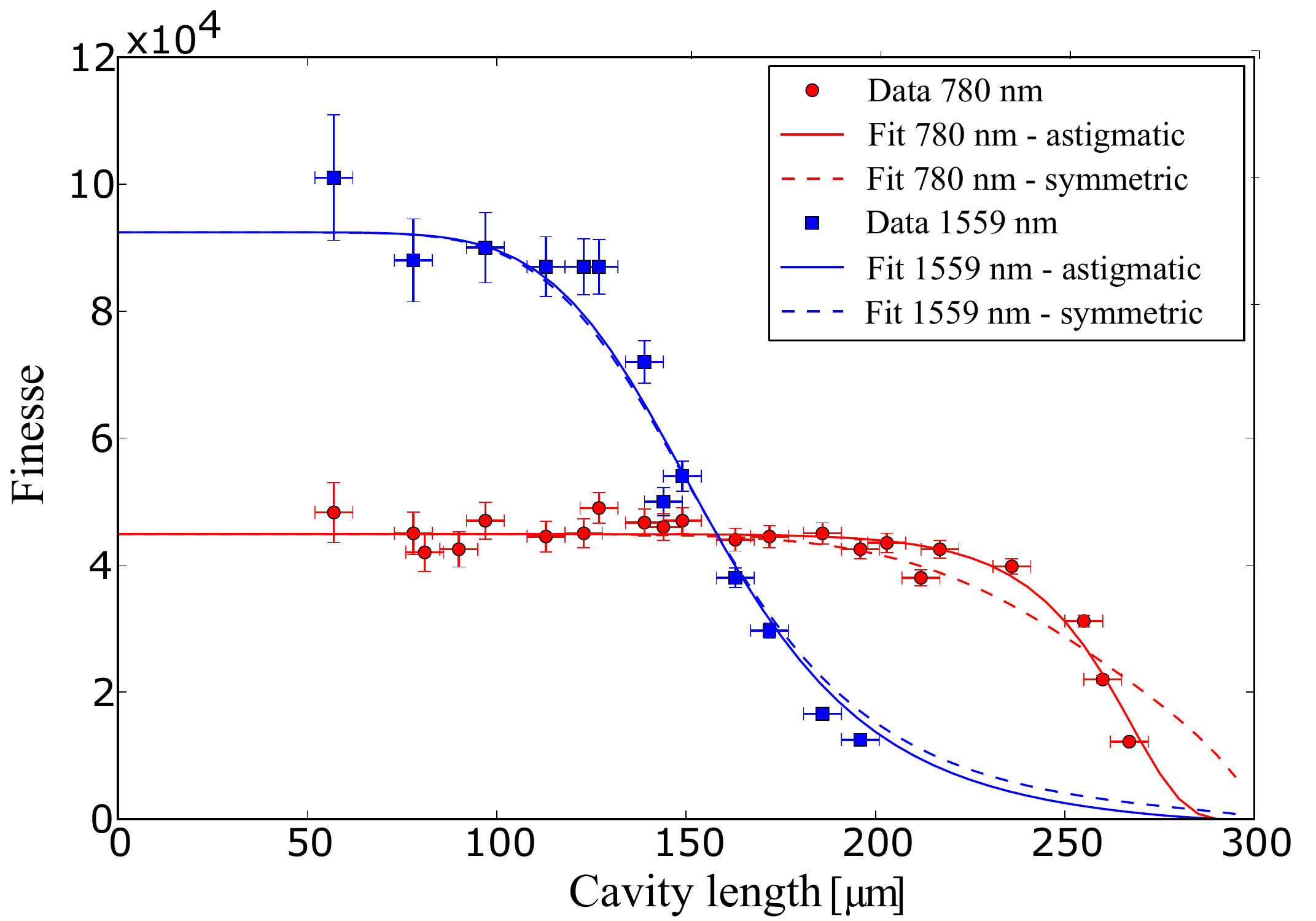}
	\caption{Finesse $\Fi$ as a function of the length, measured for 780$\,$nm (red) and 1559$\,$nm (blue) on a PC-MM cavity. The solid line is a fit using the analytical model of Eq. \ref{eq:finesse}, with only one free parameter that sets the effective size of the mirrors (see main text). The effective ROCs of the mirrors are slightly different for the two wavelengths: they are extracted from the profilometry data with a spherical fit on a region that corresponds to the mode diameter for the cavity length at which clipping losses start to dominate. 
The values of the ROCs are: 
$R_{1x}=323\,\mu$m, $R_{1y}=299\,\mu$m, $R_{2x}=290\,\mu$m, $R_{2y}=321\,\mu$m for 780$\,$nm;  
$R_{1x}=330\,\mu$m, $R_{1y}=301\,\mu$m, $R_{2x}=292\,\mu$m, $R_{2y}=323\,\mu$m for 1559$\,$nm.  The dashed line is a best-fit curve obtained by considering symmetric mirrors, \textit{i.e.} by neglecting the astigmatism of the cavity mode.}
	\label{fig:FinesseLength}
\end{figure}

To measure the resonator finesse at both wavelengths, we first align a PC-MM cavity with two slightly elliptic paraboloidal fiber mirrors. We set  the major axes to be orthogonal and measure the resonator finesse at both wavelengths. We perform this measurement by scanning the resonator length with a piezoelectric actuator and measuring the linewidth of the fundamental cavity mode at 780$\,$nm and at 1559$\,$nm in transmission. We use the sidebands generated by an electro-optic modulator for frequency calibration. A microscope is used to measure the cavity length $L$ with a precision of  $\pm5\,\mu$m, from which we calculate the free spectral-range. We perform the finesse measurement for different cavity lengths within the stability region, taking care of re-optimizing the alignment at every step. The results are shown on Fig.~\ref{fig:FinesseLength}.

We observe that, when the cavity length is shorter than 100$\,\mu$m,
the finesse stays at its maximal value of (45 $\pm$ 2)$\cdot$10$^{3}$
for 780$\,$nm and (92 $\pm$ 3)$\cdot$10$^{3}$ for 1559$\,$nm, limited
by the coating properties.  As the length increases between
100$\,\mu$m and 200$\,\mu$m, the finesse at 1559$\,$nm drops due to
the effect of the clipping losses, which start playing a major role in
reducing the finesse when the size of the mode on the mirrors grows.
For cavity length above 200$\,\mu$m, the finesse at 780$\,$nm starts
to decrease too. The mode at 780$\,$nm is less affected by the
clipping losses than the one at 1559$\,$nm due to its smaller mode
area by about a factor 2. Fabricating regular concave structures which
are large enough to avoid clipping losses for the designed cavity
length is then essential to achieve high finesse; our
CO\textsubscript{2} dot machining fabrication technique is a powerful
tool in this sense, as shown in detail in \cite{Ott2016}.

To fit the data, we use the simple analytical model described in \cite{Hunger2010}. In this model, the cavity finesse is given by 
\begin{equation}
\Fi=\frac{\pi}{\mathscr{T}+\mathscr{L}_{A}+\mathscr{L}_{S}+\left(\mathscr{L}^{M_1}_{C}+\mathscr{L}^{M_2}_{C}\right)/2}
\label{eq:finesse}
\end{equation}
where $\mathscr{T}$ is the transmission through one mirror, $\mathscr{L}_{A}$ are the absorption losses of the reflective coating, and $\mathscr{L}_{S}$ are the scattering losses due to the surface roughness of the mirror (these values stand for one mirror and we suppose them identical for the two mirrors). $\mathscr{L}^{M_i}_{C}$ are the clipping losses caused by the finite size of the mirror $M_i$.  
The value of $\mathscr{T}+\mathscr{L}_{S}+\mathscr{L}_{A}$ is determined from the finesse measurements on several short cavities, which give  (70$\pm$2)$\,$ppm at 780$\,$nm and (34$\pm$1)$\,$ppm at 1559$\,$nm. The factor close to 2 between the losses for the two wavelengths stems from the term $\mathscr{L}_{S}+\mathscr{L}_{A}$ which we estimate to be roughly five times larger at 780$\,$nm  than at 1559$\,$nm (measured value at 780$\,$nm (35$\pm$2)$\,$ppm). 

To take into account the elliptic paraboloidal structure of the mirror, we extend the previous model of the clipping losses\cite{Hunger2010, Ott2016}. We associate to each mirror an effective elliptical cross-section with principal semi-axes $a_{i}$ and $b_i$ and we take into account the elliptical and astigmatic spatial distribution of the cavity mode. The mirror's ellipse $\textrm{E}(a_{i},b_{i})$ sets the size of the cut-off for the cavity mode through the relation
\begin{equation}
\mathscr{L}^{M_i}_{C} \left(w_{i,a}, w_{i,b},a_{i},b_{i}\right) = 1-\frac{2}{\pi w_{i,a}w_{i,b}} \iint_{\textrm{E}(a_{i}, b_{i})} \textrm{exp}\left(-\frac{2x^{2}}{w^2_{i,a}}-\frac{2y^{2}}{w^2_{i,b}}\right)\textrm{d}x\textrm{d}y
\label{eq:cliploss}
\end{equation}
where $w_{i,a}, w_{i,b}$ are the $1/e^{2}$ mode radii along the principal axes of of the mirror $M_i$.

Each dataset is fitted with this model, from which we deduce the effective size of the mirrors. The mode waists on the mirrors are deduced from the cavity length and from the ROCs of the fiber mirrors.  They are slightly different for the two wavelengths. They are indeed extracted from the fibers profilometry data by using a bi-dimensional fit within a circular region, whose radius is  the mode radius at the cavity length where clipping losses become dominant (\textit{i.e.} the finesse drops to half of its maximum value). Since the fibers profile slightly deviates from an ideal elliptic paraboloid, the effective ROCs have a small dependence on the fitting region. For the measurement shown in Fig.~\ref{fig:FinesseLength}, we set the fast axes of the mirrors to be orthogonal by finding the configuration that minimizes the birefringence.

We assume that the mirror shape is an asymmetric paraboloid, truncated at $\pm a_{i}$ and $\pm b_{i}$ respectively along $x$ and $y$. This sets the constraint $\frac{a_{i}}{b_{i}}=\sqrt{\frac{R_{ix}}{R_{iy}}}$, where $R_{ix,y}$ are the ROCs from the profilometry. Since our cavity is only slighlty asymmetric, it is difficult to estimate $a_{i}$ and $b_{i}$ independently with good confidence because, as remarked in \cite{Ott2016}, larger clipping loss on one mirror can be compensated by smaller losses on the other. We thus impose the additional constraint $a_{1}=a_{2}\sqrt{\frac{R_{1x}}{R_{2x}}}$, \textit{i.e.} that the cutoff occurs at the same depth on both mirrors. 

The best fit curves give us a value of about 20$\,\mu$m for the semi-axes of the elliptical cross-section describing the effective mirror size, with a difference of 1$\,\mu$m between the major and minor axes.
We also find a dependence of the effective size on the wavelength:  the ones at 1559$\,$nm  exceed by about 2$\,\mu$m the ones at 780$\,$nm.
At this distance of 20$\,\mu$m from the center of the mirror, we notice on the profilometry data a deviation of about $200\,$nm from the ideal parabolic profile, that rapidly increases further away. 
 
We find good agreement between the data and the model within the experimental uncertainties, which indicates that no other main sources of loss are significantly affecting the cavity. For comparison, dashed lines in Fig.~\ref{fig:FinesseLength} show the results obtained by fitting our data by considering symmetric mirrors with a ROC equal to the average of the ROCs along the two principal axes. This corresponds to the situation where we neglect the asymmetry of the cavity mode, which becomes particularly important close to the instability region. We observe that such a model fits badly with our data at 780$\,$nm (the level of statistical significance of the best-fit is less than 0.1\%, in contrast to the 25\% significance level that we obtain with the elliptical model) \cite{Taylor1997}. This wavelength is indeed more sensitive, as clipping losses starts to play a significant role closer to the instability region where the mode size on the mirror has a sharper variation with the cavity length. The effect of the astigmatism of the mode cannot be neglected especially in the situation where the two major-axes of the elliptical cross-section are orthogonal, which is the configuration which minimize the frequency splitting of the polarization eigenmodes.

\section{Large birefringent frequency splitting with asymmetric mirrors}

Knowing the finesse of the cavity at both wavelengths, we can now turn to the measurement of the birefringent frequency splitting of the polarization eigenmodes at 780$\,$nm and at 1559$\,$nm,  and verify if both sets of data can be explained using the geometrical birefringence of the cavity. In order to achieve a large frequency splitting, we use mirror shapes, as the one presented in Fig.~\ref{fig:ellip_circ_combined}(b), which have radii of curvature of about 350$\, \mu$m and 250$\, \mu$m along their eigenaxes. 

\begin{figure}[htbp]
	\centering\includegraphics[width=0.9\columnwidth]{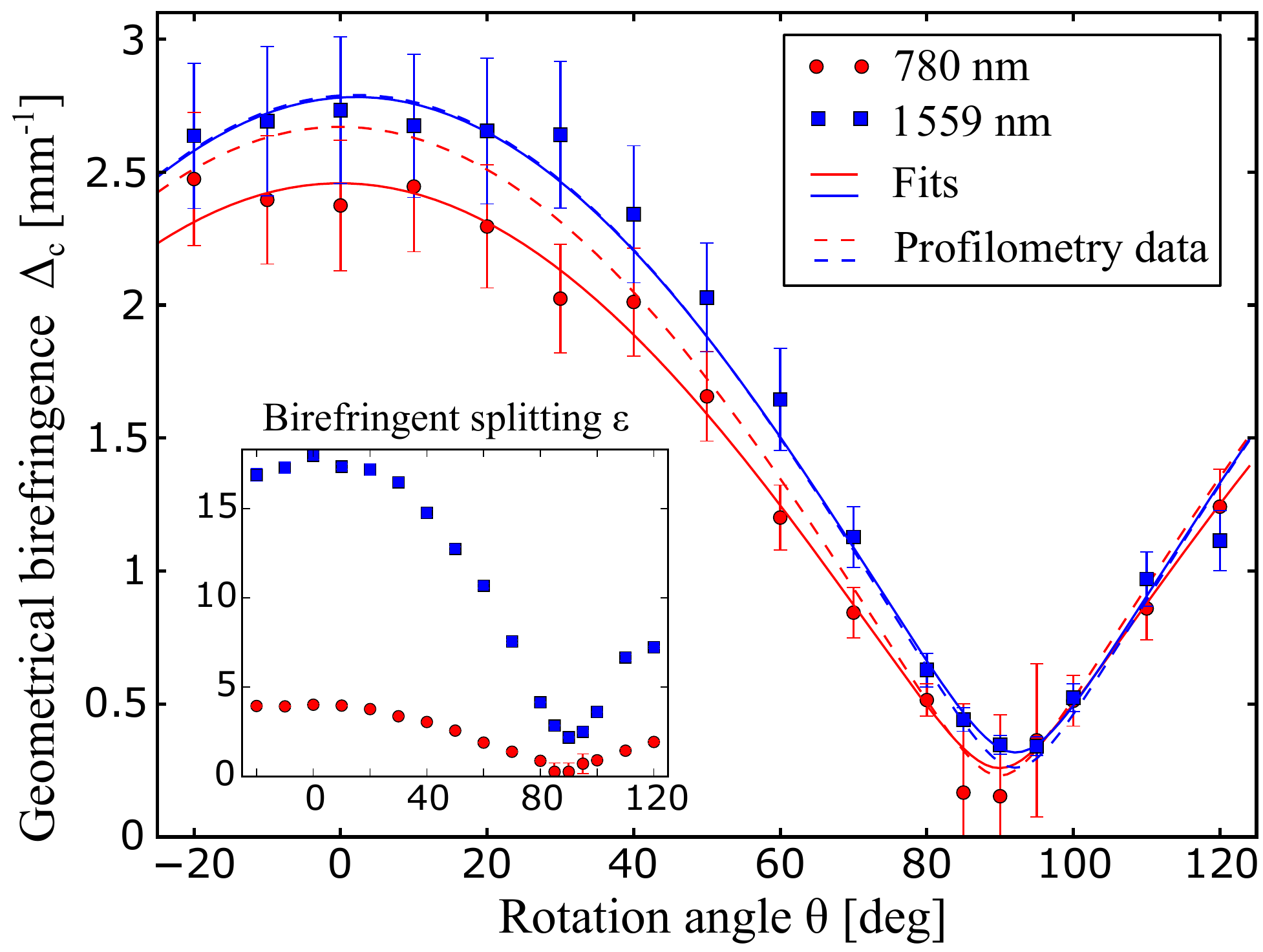}
	\caption{Geometrical birefringence retrieved from (a) after normalization by finesse and wavelength. Experimental error bars mainly come from cavity length uncertainty that implies finesse uncertainty. Full lines are fits with the geometrical birefringence of each mirror as parameters. Inset: Birefringent frequency splitting normalized by the cavity linewidth as function of the rotation angle of one fiber mirror obtained in a FFP cavity with asymmetric mirrors at 780 nm (red dots) and 1559 nm (blue squares).}
	\label{fig:Birefringence}
\end{figure}

To measure the birefringent frequency splitting, we align a FFP cavity with two SM fibers. The resonator length is set to $L \simeq \left( 55\pm 5 \right) \,\mu$m and we rotate one of the fiber around its axis. The relatively short length facilitates the required re-optimisation of the cavity alignment after a rotation step. We observe the evolution of the absolute birefringent frequency splitting and the cavity linewidth for the two wavelengths. The birefringent frequency splitting $\epsilon$ data is shown on the inset of the Figure~\ref{fig:Birefringence}. The difference by approximately a factor 4 in the splitting between the two data sets stems from the wavelength dependence of Eq. \ref{eq:splitting} and from the higher finesse at 1559$\,$nm ($\Fi_{1559} \simeq 90\,000 $) than at 780$\,$nm ($\Fi_{780} \simeq 45\,000 $). 

Figure~\ref{fig:Birefringence} shows the geometrical birefringence $\Delta_c$ obtained by dividing the birefringent frequency splitting by the wavelength, the finesse and $2 \pi^2$ (see Eq. \ref{eq:splitting}). The fact that the two sets of data are relatively close to each other confirms that the birefringent frequency splitting originates from the geometrical asymmetry of the mirrors. 

The small remaining mismatch can be explained by the slight differences in the effective ROCs seen by the two wavelengths. Indeed, the effective area of the mirror seen by the cavity mode is twice larger at 1559$\,$nm than at 780$\,$nm.  
From the profilometry measurement, we calculate the geometrical birefringence of each mirror for the two wavelengths: 
$\left[\delta_1= 1.45(2) \, \mathrm{mm}^{-1},~\delta_2=  1.22(2)\, \mathrm{mm}^{-1}\right]$ at 780$\,$nm and 
$\left[\delta_1=  1.53(1) \, \mathrm{mm}^{-1},~\delta_2= 1.27(1) \, \mathrm{mm}^{-1}\right]$ at 1559$\,$nm, where the number in parentheses is the numerical value of 
standard uncertainty referred to the corresponding last digits of the quoted result.
We can then deduce the expected cavity geometrical birefringences for the two wavelengths, which are plotted in dotted lines. When fitting with the geometrical birefringence of each mirror as parameters, we obtain a good matching of fitted curves (full lines) with the data points for the following values   
$\left[ \delta_{1,\mathrm{f}}=  1.36(5) \, \mathrm{mm}^{-1},~\delta_{2,\mathrm{f}}=  1.11(7) \, \mathrm{mm}^{-1}\right]$ at 780$\,$nm and 
$\left[ \delta_{1,\mathrm{f}}=  1.55(4) \, \mathrm{mm}^{-1},~\delta_{2,\mathrm{f}}=  1.23(5) \, \mathrm{mm}^{-1}\right]$ at 1559$\,$nm. The fitted geometric birefringences at 780$\,$nm are slightly lower than the predicted values, which correspond to the small shift of the data points to lower values. We attribute the remaining mismatch to slight differences between the effective ROCs actually experienced by the cavity mode and the ones given by the fit of profilometry data in the center of the cavity. Indeed, the actual position of the cavity mode can deviate slightly from the center of the fiber while maintaining a good finesse and coupling, which were the parameters we optimized during the alignment. Thus, slight deviations of the profile from an ideal paraboloid and small variations of the mode position on the mirror lead to small differences in the effective ROCs. Nevertheless, all data points lie within a $3$\% uncertainty range on the ROCs given by the profilometry measurement. 
This confirms that the geometric birefringence of the mirrors extracted from in-situ profilometry can be used to predict the actual geometrical birefringence of the cavity, which is the key parameter that controls the birefringent splitting for a given finesse and wavelength. Our laser ablation setup with multiple pulses and profilometry allows us to realize and characterize fiber mirror shapes with sufficient precision to reach such a control and be able to tailor the birefringence properties of fiber cavities.

\section{Conclusion} 

We have demonstrated how the laser dot machining method allows
controlling the birefringence of fiber Fabry-Perot cavities. By
defining the geometrical birefringence, we quantify the intrinsic
geometric property of a mirror that leads to frequency
splitting of the polarization eigenmodes. We discussed the polarization and spatial modes of the cavity in the case of asymmetric mirrors.  Our laser ablation
technique with multiple pulses in a precise pattern has been used to
produce elliptic paraboloidal shapes on the fiber
endfacets. Elliptical pulse patterns lead to strongly asymmetrical
shapes with large geometric birefringences. The same method can also
be used to obtain cavities with very low geometric birefringence by
producing mirror shapes with very small asymmetry without rotation of
the substrate thanks to optimized pulses patterns.  We have realized
fiber cavities that are resonant in two frequency bands separated by
an octave with a specific dual-wavelength coating deposited on the
machined fibers. High-finesse operation can be achieved with SM-SM and PCF-MM
fiber cavities that ensure stable injection at both wavelengths.  In the context of elliptic paraboloidal
mirrors, the astigmatic spatial distribution of the mode needs to be
taken into account to describe the clipping losses limit of long
cavities.  We observed geometric birefringences of a cavity at both
wavelengths with identical values within the measurement precision,
and consistent with the profilometry data. This result directly
confirms the geometric origin of the birefringent frequency splitting of
fiber Fabry-Perot cavities.

The precise control of the polarization opens new opportunities for
fiber cavity applications, especially for cavity QED experiments,
allowing either large frequency polarization eigenmodes splitting or degenerate mode
configuration. In addition to the polarization, non-symmetric
structures carved into a substrate can also be used to tailor the
spatial distribution of the cavity mode. The extension of the laser
dot machining technique to produce shapes beyond the simple rotational
symmetry also constitutes a first step towards realization of more
complex microscopic structures to engineer new types of stable optical
resonators.

\section*{Funding}
This work was supported by the Agence Nationale de la Recherche (ANR)
(SAROCEMA project, ANR-14-CE32-0002) and the
European Research Council (ERC) (Advanced Grant ``EQUEMI'', GA 671133).

\section*{Acknowledgments}
We thank K.~Sch\"{u}ppert for contributions in the early stage of the experiment.
While finalizing the manuscript, we become aware of a related method of producing asymmetric mirrors \cite{Cui2018}.


\begin{thebibliography}{10}
\newcommand{\enquote}[1]{``#1''}

\bibitem{Colombe2007}
Y.~Colombe, T.~Steinmetz, G.~Dubois, F.~Linke, D.~Hunger, and J.~Reichel,
  \enquote{{Strong atom-field coupling for Bose-Einstein condensates in an
  optical cavity on a chip},} Nature \textbf{450}, 272--276 (2007).

\bibitem{Hunger2010}
D.~Hunger, T.~Steinmetz, Y.~Colombe, C.~Deutsch, T.~W. H{\"{a}}nsch, and
  J.~Reichel, \enquote{{A fiber Fabry-Perot cavity with high finesse},} New
  Journal of Physics \textbf{12}, 065038 (2010).

\bibitem{Barontini2015}
G.~Barontini, L.~Hohmann, F.~Haas, J.~Esteve, and J.~Reichel,
  \enquote{{Deterministic generation of multiparticle entanglement by quantum
  Zeno dynamics},} Science \textbf{349}, 1317--1321 (2015).

\bibitem{Uphoff2015}
M.~Uphoff, M.~Brekenfeld, G.~Rempe, and S.~Ritter, \enquote{{Frequency
  splitting of polarization eigenmodes in microscopic Fabry-Perot cavities},}
  New Journal of Physics \textbf{17}, 013053 (2015).

\bibitem{Gallego2016}
J.~Gallego, S.~Ghosh, S.~K. Alavi, W.~Alt, M.~Martinez-Dorantes, D.~Meschede,
  and L.~Ratschbacher, \enquote{{High-finesse fiber Fabry-Perot cavities:
  stabilization and mode matching analysis},} Applied Physics B \textbf{122},
  47 (2016).

\bibitem{Brandstatter2013}
B.~Brandst{\"{a}}tter, A.~McClung, K.~Sch{\"{u}}ppert, B.~Casabone, K.~Friebe,
  A.~Stute, P.~O. Schmidt, C.~Deutsch, J.~Reichel, R.~Blatt, and T.~E. Northup,
  \enquote{{Integrated fiber-mirror ion trap for strong ion-cavity coupling},}
  Review of Scientific Instruments \textbf{84}, 123104 (2013).

\bibitem{Steiner2013}
M.~Steiner, H.~M. Meyer, C.~Deutsch, J.~Reichel, and M.~K{\"{o}}hl,
  \enquote{{Single ion coupled to an optical fiber cavity},} Physical Review
  Letters \textbf{110}, 043003 (2013).

\bibitem{Toninelli2010}
C.~Toninelli, Y.~Delley, T.~St{\"{o}}ferle, A.~Renn, S.~G{\"{o}}tzinger, and
  V.~Sandoghdar, \enquote{{A scanning microcavity for in situ control of
  single-molecule emission},} Applied Physics Letters \textbf{97}, 021107
  (2010).

\bibitem{Muller2010}
A.~Muller, E.~B. Flagg, J.~R. Lawall, and G.~S. Solomon,
  \enquote{{Ultrahigh-finesse, low-mode-volume Fabry-Perot microcavity.}}
  Optics letters \textbf{35}, 2293--5 (2010).

\bibitem{Miguel-Sanchez2013}
J.~Miguel-S{\'{a}}nchez, A.~Reinhard, E.~Togan, T.~Volz, A.~Imamoglu, B.~Besga,
  J.~Reichel, and J.~Est{\`{e}}ve, \enquote{{Cavity quantum electrodynamics
  with charge-controlled quantum dots coupled to a fiber Fabry-Perot cavity},}
  New Journal of Physics \textbf{15}, 045002 (2013).

\bibitem{Albrecht2013}
R.~Albrecht, A.~Bommer, C.~Deutsch, J.~Reichel, and C.~Becher,
  \enquote{{Coupling of a single nitrogen-vacancy center in diamond to a
  fiber-based microcavity},} Physical Review Letters \textbf{110}, 243602
  (2013).

\bibitem{Kaupp2016}
H.~Kaupp, T.~H{\"{u}}mmer, M.~Mader, B.~Schlederer, J.~Benedikter, P.~Haeusser,
  H.~C. Chang, H.~Fedder, T.~W. H{\"{a}}nsch, and D.~Hunger,
  \enquote{{Purcell-Enhanced Single-Photon Emission from Nitrogen-Vacancy
  Centers Coupled to a Tunable Microcavity},} Physical Review Applied
  \textbf{6}, 054010 (2016).

\bibitem{Jeantet2016}
A.~Jeantet, Y.~Chassagneux, C.~Raynaud, P.~Roussignol, J.~S. Lauret, B.~Besga,
  J.~Est{\`{e}}ve, J.~Reichel, and C.~Voisin, \enquote{{Widely Tunable
  Single-Photon Source from a Carbon Nanotube in the Purcell Regime},} Physical
  Review Letters \textbf{116}, 1--5 (2016).

\bibitem{Hummer2016}
T.~H{\"{u}}mmer, J.~Noe, M.~S. Hofmann, T.~W. H{\"{a}}nsch, A.~H{\"{o}}gele,
  and D.~Hunger, \enquote{{Cavity-enhanced Raman Microscopy of Individual
  Carbon Nanotubes},} Nature Communications \textbf{7}, 1--7 (2016).

\bibitem{Flowers-Jacobs2012}
N.~E. Flowers-Jacobs, S.~W. Hoch, J.~C. Sankey, A.~Kashkanova, A.~M. Jayich,
  C.~Deutsch, J.~Reichel, and J.~G.~E. Harris, \enquote{{Fiber-cavity-based
  optomechanical device},} Applied Physics Letters \textbf{101}, 221109 (2012).

\bibitem{Kashkanova2016}
A.~D. Kashkanova, A.~B. Shkarin, C.~D. Brown, N.~E. Flowers-Jacobs,
  L.~Childress, S.~W. Hoch, L.~Hohmann, K.~Ott, J.~Reichel, and J.~G.~E.
  Harris, \enquote{{Superfluid Brillouin optomechanics},} Nature Physics
  \textbf{13}, 74--79 (2016).

\bibitem{Zhong2017}
H.~Zhong, G.~Fl{\"{a}}schner, A.~Schwarz, R.~Wiesendanger, P.~Christoph,
  T.~Wagner, A.~Bick, C.~Staarmann, B.~Abeln, K.~Sengstock, and C.~Becker,
  \enquote{{A millikelvin all-fiber cavity optomechanical apparatus for merging
  with ultra-cold atoms in a hybrid quantum system},} Review of Scientific
  Instruments \textbf{88}, 023115 (2017).

\bibitem{Barbour2011}
R.~J. Barbour, P.~A. Dalgarno, A.~Curran, K.~M. Nowak, H.~J. Baker, D.~R. Hall,
  N.~G. Stoltz, P.~M. Petroff, and R.~J. Warburton, \enquote{A tunable
  microcavity,} J.~Appl.~Phys. \textbf{110}, 053107 (2011).

\bibitem{Petrak2011}
B.~Petrak, K.~Konthasinghe, S.~Perez, and A.~Muller,
  \enquote{Feedback-controlled laser fabrication of micromirror substrates,}
  Rev.~Sci.~Instrum. \textbf{82}, 123112 (2011).

\bibitem{Hunger2012}
D.~Hunger, C.~Deutsch, R.~J. Barbour, R.~J. Warburton, and J.~Reichel,
  \enquote{Laser micro-fabrication of concave, low-roughness features in
  silica,} AIP Adv. \textbf{2}, 012119 (2012).

\bibitem{Greuther2014}
L.~Greuther, S.~Starosielec, D.~Najer, A.~Ludwig, L.~Duempelmann, D.~Rohner,
  and R.~J. Warburton, \enquote{A small mode volume tunable microcavity:
  Development and characterization,} Appl.~Phys.~Lett. \textbf{105}, 121105
  (2014).

\bibitem{Ott2016}
K.~Ott, S.~Garcia, R.~Kohlhaas, K.~Sch{\"{u}}ppert, P.~Rosenbusch, R.~Long, and
  J.~Reichel, \enquote{{Millimeter-long fiber Fabry-Perot cavities},} Optics
  Express \textbf{24}, 9839 (2016).

\bibitem{Bielsa2009}
F.~Bielsa, A.~Dupays, M.~Fouch{\'e}, R.~Battesti, C.~Robilliard, and C.~Rizzo,
  \enquote{Birefringence of interferential mirrors at normal incidence,}
  Appl.~Phys.~B \textbf{97}, 457 (2009).

\bibitem{Fleisher2016}
A.~J. Fleisher, D.~A. Long, Q.~Liu, and J.~T. Hodges, \enquote{Precision
  interferometric measurements of mirror birefringence in high-finesse optical
  resonators,} Phys. Rev. A \textbf{93}, 013833 (2016).

\bibitem{Stute2012}
A.~Stute, B.~Casabone, P.~Schindler, T.~Monz, P.~O. Schmidt,
  B.~Brandst{\"{a}}tter, T.~E. Northup, and R.~Blatt, \enquote{{Tunable
  ion-photon entanglement in an optical cavity},} Nature \textbf{485}, 482--485
  (2012).

\bibitem{Reiserer2014}
A.~Reiserer, N.~Kalb, G.~Rempe, and S.~Ritter, \enquote{{A quantum gate between
  a flying optical photon and a single trapped atom.}} Nature \textbf{508},
  237--40 (2014).

\bibitem{Takahashi2014}
H.~Takahashi, J.~Morphew, F.~Oru{\v{c}}evi{\'{c}}, A.~Noguchi, E.~Kassa, and
  M.~Keller, \enquote{{Novel laser machining of optical fibers for long
  cavities with low birefringence.}} Optics express \textbf{22}, 31317--28
  (2014).

\bibitem{Gehr2010}
R.~Gehr, J.~Volz, G.~Dubois, T.~Steinmetz, Y.~Colombe, B.~L. Lev, R.~Long,
  J.~Est{\`{e}}ve, and J.~Reichel, \enquote{{Cavity-based single atom
  preparation and high-fidelity hyperfine state readout},} Physical Review
  Letters \textbf{104} (2010).

\bibitem{Volz2011}
J.~Volz, R.~Gehr, G.~Dubois, J.~Est{\`{e}}ve, and J.~Reichel,
  \enquote{{Measurement of the internal state of a single atom without energy
  exchange.}} Nature \textbf{475}, 210--3 (2011).

\bibitem{jones1941new}
R.~C. Jones, \enquote{A new calculus for the treatment of optical systems i.
  description and discussion of the calculus,} J. Opt. Soc. Am. \textbf{31},
  488--493 (1941).

\bibitem{Arnaud1969}
J.~A. Arnaud and H.~Kogelnik, \enquote{{Gaussian Light Beams with General
  Astigmatism},} Applied Optics \textbf{8}, 1687 (1969).

\bibitem{Habraken2007}
S.~J. Habraken and G.~Nienhuis, \enquote{{Modes of a twisted optical cavity},}
  Physical Review A - Atomic, Molecular, and Optical Physics \textbf{75}, 1--11
  (2007).

\bibitem{Weber2012}
H.~Weber, \enquote{{Rays and fields in general astigmatic resonators},} Journal
  of Modern Optics \textbf{59}, 740--770 (2012).

\bibitem{Taylor1997}
J.~R. Taylor, \emph{An introduction to error analysis: the study of
  uncertainties in physical measurements} (University Science Books, Sausalito,
  California, 1997), 2nd ed.

\bibitem{Cui2018}
J.-M. Cui, K.~Zhou, M.-S. Zhao, M.-Z. Ai, C.-K. Hu, Q.~Li, B.-H. Liu, J.-L.
  Peng, Y.-F. Huang, C.-F. Li, and G.-C. Guo, \enquote{{Polarization
  nondegenerate fiber Fabry-Perot cavities with large tunable splittings},}
  Applied Physics Letters \textbf{112}, 171105 (2018).

\end{thebibliography}
\end{document}